# Phase Resolved Observations of Temporal Soliton Pulse Propagation in Silicon Nanowires

Matthew Marko [1, 2, *], Xiujian Li [2, 3], Jiangjun Zheng [2], Jiali Liao [3], Chad Husko [4],

[1] Navy Air Warfare Center Aircraft Division (NAWCAD), Joint Base McGuire-Dix-Lakehurst, Lakehurst NJ 08733, USA
[2] Optical Nanostructures Laboratory, Columbia University in the City of New York, New York NY 10027, USA
[3] Tech-Physical Research Center, Science College, National University of Defense Technology, Changsha, Hunan 410073, China
[4] Center for Ultrahigh bandwidth Devices for Optical Systems (CUDOS), School of Physics, University of Sydney, Australia
matthew.marko@navy.mil [*]

## I. Abstract

An effort was conducted to study temporal soliton pulse propagation in silicon nano-waveguides. These nonlinear phenomenas were studied both numerically and experimentally with phase-resolved Frequency Resolved Optical Gating. Soliton pulse broadening, as well as pulse splitting from two-photon absorption, was observed experimentally, and the simulations matched all of the experimental results. Further simulations with the validated model have demonstrated that compression can be observed in centimeter-length waveguides. This study has demonstrated the feasibility of self-sustaining soliton pulse propagation at substantially shorter length scales than optical fibers, which offers much potential applications with regards to all-optical data transfer and computing.

## II. Introduction

Optical solitons are undistorted standing waves that can propagate over long distances undistorted, which result as part of the interplay between the nonlinear Kerr effect, and anomalous dispersion [1]. When there is strong SPM interplaying with strong anomalous group velocity dispersion (GVD), then soliton pulse compression may occur [2]. The SPM will cause the leading edge frequencies to be lowered, and the trailing edge frequencies to be raised. At the same time, the anomalous GVD will cause the lowered leading edge frequencies to slow down, and the raised trailing edge frequencies to speed up. This results in the pulse narrowing, or compressing into what is known as the soliton pulse.

Soliton pulses within fiber optics is a well-established subject, and it has been used previously for data communication over many kilometers of fiber networks. The primary disadvantage of fibers is simply the long-length scales required for the optical nonlinearities to occur. There would be many advantages to having soliton pulse compression occur at the centimeter or millimeter length scales, which would allow for better optical data transfer and processing within a typical computer chip.

Temporal pulse compression has been observed in silicon waveguides through the use of autocorrelation [3] and cross-correlation [4]. While autocorrelation and cross-correlation can provide an approximation of the temporal pulse duration, it provides little information regarding the pulse shape and phase [5,6]. In fact, autocorrelation can sometime provide misleading data in the case of pulse splitting or unusual pulse-shaping, as autocorrelation requires assuming a perfect pulse-shape function. For this reason, the authors feel it is necessary to observe temporally the input and output of silicon nanowires with the use of Frequency Resolved Optical Gating, which offers an accurate shape of the pulse intensity and phase, both at the input and output of the waveguide.

One of the biggest challenges to soliton pulse compression is the nonlinear effects of multi-photon and free-carrier absorption [7,8]. All dielectric materials have an electronic band gap [9], where photons with energy greater than the electronic band gap get absorbed by the material. The electronic band gap for silicon is 1.1 electron-volts (eV). As it is desired to use light pulses at a wavelength of 1550 nm, this corresponds to a photon energy of 0.8 eV, and thus silicon is transparent at this wavelength. However, multi-photon absorption is the result of multiple photons interacting, and their combined energy exceeding the electronic band gap. At 1550 nm, two photons will result in a total energy of 0.8*2=1.6 eV, which exceeds the 1.1 eV electronic band gap of silicon, and therefore, silicon is subject to two-photon absorption at 1550 nm.

**III. Numerical Modeling of Silicon Nano-Waveguides with Two-Photon Absorption**

A split-step NLSE code [1,10-12] was written by the authors in an effort to demonstrate the challenges of achieving soliton pulse compression in a silicon nano-waveguide. A model was created with MatLab to replicate an ultrashort pulse propagating through a 450*250 nm silicon channel waveguide. The dispersion parameters were found numerically [13], and a value of 4.5 $ps^2$/m was used consistently for the channel waveguide. The model was developed to determine the coefficient of TPA [14,15], and attenuate the pulse based on the predicted TPA and FCA [16,17]. The physical length of the waveguide was set at 10 cm in order to better observe the effect of soliton formation over longer distances, which are characterized by their self-sustaining nature over long propagation distances.

The model was run by propagating an ideal hyperbolic-secant squared pulse with no chirp through the 10 cm long waveguide, with the known parameters for dispersion, self-phase modulation, and two-photon absorption. With an input pulse energy of 100 fJ, the pulse shape looked almost identical, save for the effects of linear loss. With an input pulse energy of 1 pJ, minor temporal compression was observed on the output of the waveguide. With an increase in pulse energy of 10 pJ, even in the presence of two-photon and free-carrier absorption, soliton pulse compression was clearly observed at the output of the waveguide, with a compression factor ($T_{FWHM}$ input / $T_{FWHM}$ output) of 7. This numerical modeling demonstrated the feasibility of temporal soliton pulse compression in a 10 centimeter silicon nano-waveguide, a far shorter length than typically observed in optical fibers.

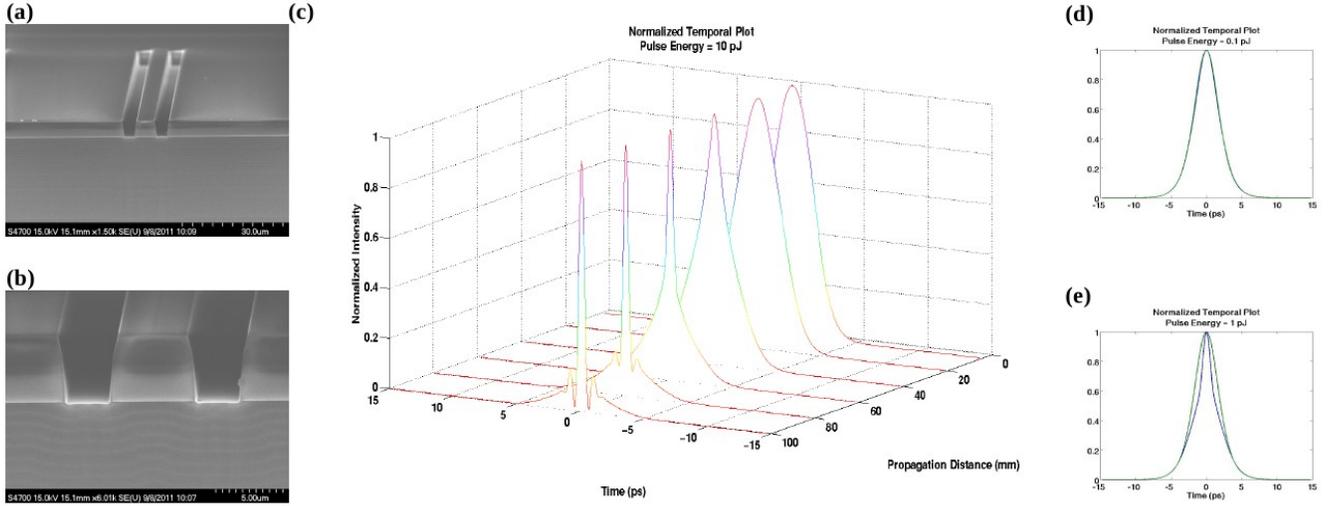

FIG. 1. Waveguide Facets under Scanning Electron Microscope (SEM) for units of (a) 30 μm and (b) 5 μm. Waterfall plot (c) of NLSE simulations of soliton pulse compression for a 10 pJ pulse in a 10 cm waveguide. NLSE simulations output after 10 cm pulse propagation of a (d) 100 fJ and (e) 1 pJ pulse.

One of the challenges with regards to soliton pulse compression in silicon waveguides is the long lengths relative to most chip-scale applications. For the 2.3 ps pulse used in this simulation, the nonlinear length $L_{NL} = |\beta_2| / \tau^2$ is over 85 cm long; much longer than the waveguide used in the simulations. In order to observe this soliton pulse compression at these length scales, it was necessary to use relatively high soliton numbers N = (Pulse Energy / Fundamental Soliton Energy)^½, where the fundamental soliton energy is [1,2]:

$$E_p = \frac{2|\beta_2|}{|\gamma|\tau} \tag{1}$$

$$\gamma = \frac{2\pi n_2}{Area_{effective} \lambda} \cdot \left(\frac{n_{group}}{n_{effective}}\right)^2 \tag{2}$$

where γ is the nonlinear coefficient, $n_2$ is the 3rd order optical Kerr coefficient, λ is the wavelength, $\beta_2$ is the second order GVD coefficient, and τ is the temporal width of the input pulse. In this particular study, the 100 fJ, 1 pJ, and 10 pJ pulses had soliton numbers of 3.63, 11.48, and 36.30, respectively. As all of these pulses have energies greater than that of the fundamental soliton, temporal compression would eventually occur at a long enough waveguide length. Since the overall goal of this project is a self-sustaining pulse, however, compression itself is not necessary in practical applications. The goal of using these pulses is to ensure that the dispersion will not compromise the pulse, an easy objective considering the low fundamental soliton energies within these silicon waveguides.

**IV. Experimental Studies of Pulse Propagation through Silicon Nanowire-Waveguides**

An experimental effort was undertaken to observe directly the temporal effects after nonlinear optical propagation within a silicon nanowire. Silicon nano-waveguides were fabricated by the Institute of Microelectronics (IME) of Singapore. These silicon-on-insulator waveguides were designed to have interior dimensions of 250 nm by 450 nm, and a linear length of 4 mm. The next step was to couple the silicon waveguide. Free-space coupling with objective lenses were used, utilizing a half-wave plate and a linear polarizer in order to ensure that the light entered the waveguide with consistent TE polarization.

The next step was to set up an apparatus for Frequency Resolved Optical Gating (FROG) autocorrelation [5,6]. As modern sensors are physically unable to detect anything substantially shorter than a nanosecond, it is necessary to use autocorrelation to study the temporal effects of these pulses. Autocorrelation involves using interferometry, where the output is a function of the delay, or shifting of one of the beams after the pulse has been sent through a beam-splitter. A commercial autocorrelator was first attempted, but this was abandoned as the autocorrelator was limited to simple pulse duration measurements; it is much more desirable to measure the actual temporal shape and phase of the pulse.

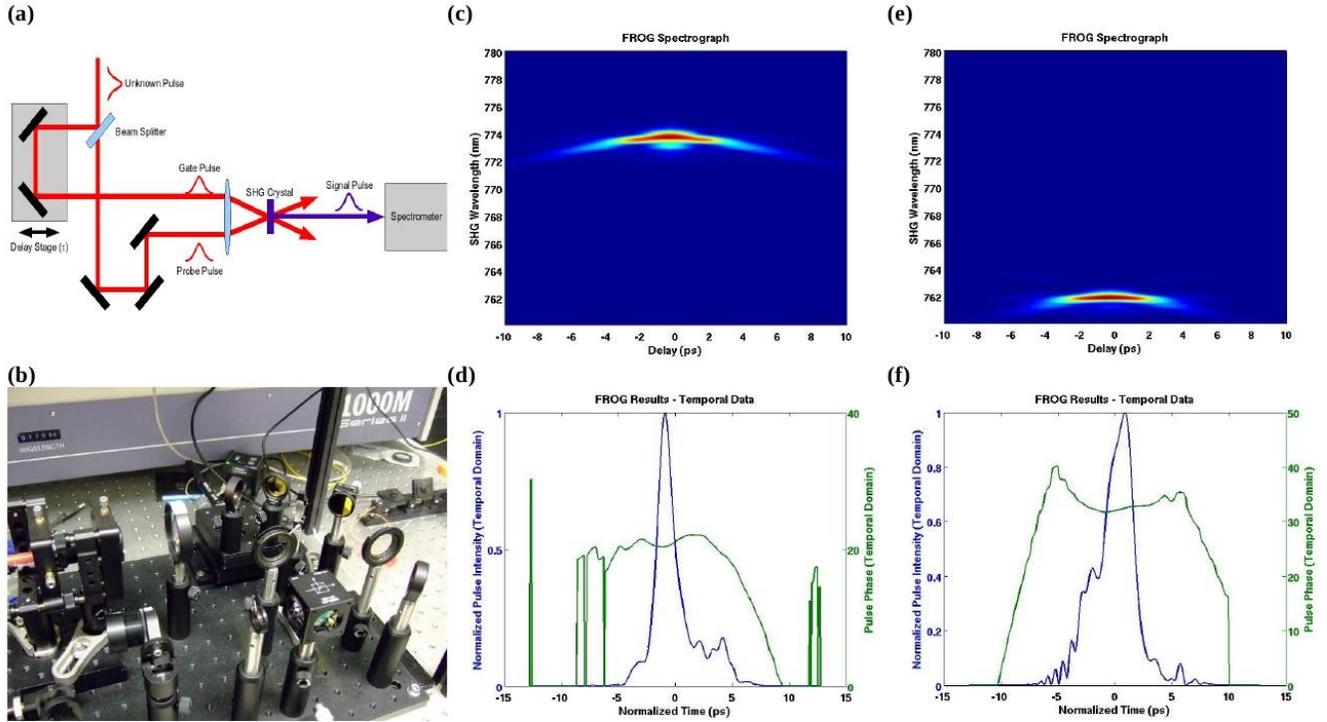

FIG. 2. Diagram (a) of FROG apparatus, and photograph of FROG set-up used in this experiment. FROG spectrogram (c) and temporal pulse magnitude and phase (d) of mode-locked laser at λ = 1560 nm. FROG spectrogram (e) and temporal pulse magnitude and phase (f) of mode-locked laser at λ = 1536 nm.

After setting up the FROG, the silicon device was coupled to the laser in order to attempt to measure the self-phase modulation and soliton compression. In order to achieve sufficient sensitivity to detect the attenuated pulses, the FROG autocorrelator was coupled to a spectrometer utilizing a liquid-nitrogen-cooled CCD detector. This new spectrometer was able to detect pulsed signals down to an average power of -25 dBm, thus giving the FROG a sensitivity (Power $_{peak}$*Power $_{average}$) of 0.18 mW$^2$. With this improved sensitivity, the FROG was able to directly detect the SHG of the pulses after being attenuated through the silicon waveguide, without requiring the use of an Erbium Doped Fiber Amplified (EDFA) on the output. This capability has provided true phase-resolved temporal measurements of the silicon waveguide, without having to question the impacts of nonlinearities from the EDFA.

The input source was a tunable mode-locked fiber laser, with a repetition rate of 39.11 MHz, and an average measured pulse energy of 600 pJ. The experiment was run at different wavelengths ranging from 1536 to 1560 nm. The input pulse's polarization was controlled with a fiber polarization controller before being sent through free-space, and a linear polarizer was used to ensure consistent TE polarization at the input. In between the fiber and the free-space polarizer, a half-wave plate was used in order to control the pulse energy while preserving the pulse shape and phase at the different energy levels, as nonlinearities within the mode-locked laser have been observed at different pump settings. The spectral results of FROG measurements of the mode-locked laser have been collected and compared with measurements from an optical spectrum analyzer, demonstrating that the FROG is properly aligned.

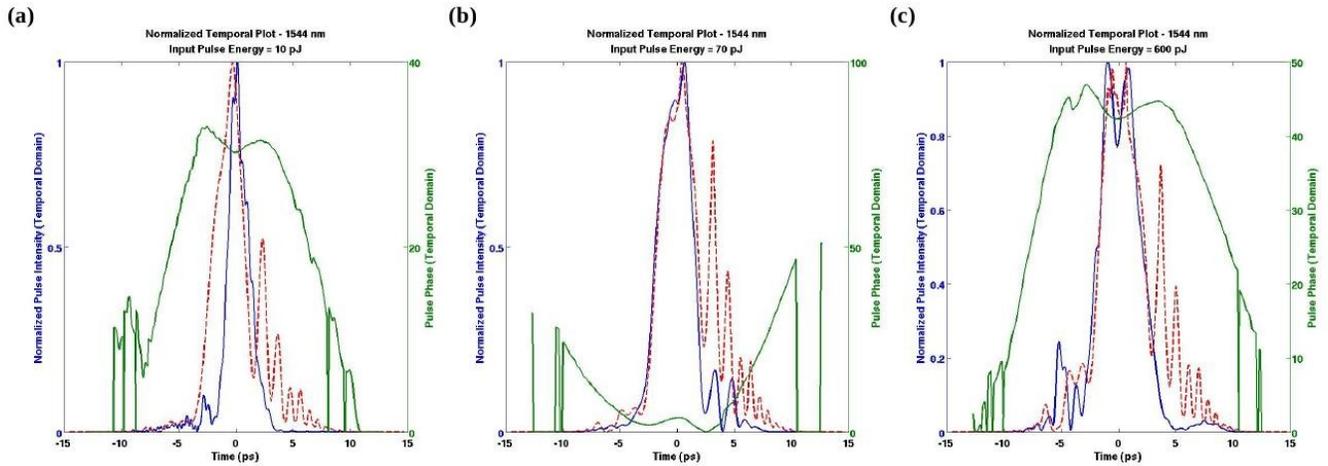

FIG. 3. Comparisons between NLSE and FROG experimental results. Blue is the experimental FROG pulse magnitude, green is the experimental FROG pulse-phase, and dashed-red is the NLSE simulation of the pulse magnitude, for (a) 10 pJ, (b) 70 pJ, and (c) 600 pJ (excluding coupling loss).

At the various different pulse energies, the FROG experimental results closely match the same NLSE simulation results, for the specific 4 mm waveguide used in the experiment. This NLSE model was identical to the model used in the earlier study of soliton pulse compression, with two exceptions: the length of the waveguide was set at 4.1 mm; and the input pulse-shape used was the FROG measured laser pulse input, rather than an ideal hyperbolic secant squared pulse.

In the low-energy measurement, with a 10 pJ (before any coupling loss) pulse, there was very little change in the pulse after just 4 mm. With the medium energy measurement of 70 pJ (before any coupling loss), there was clear evidence of pulse broadening, and one could fairly make out signs of the pulse starting to split. One clear distinction between the FROG measurements and the NLSE simulations was noticed with the presence of noise and side-lobes apart from the pulse. In the NLSE simulations, the noise was amplified, whereas in the experimental traces, the noise was suppressed; this is believed to be the result of limitations on the sensitivity of the spectrometer.

More profound nonlinearity was observed at a pulse energy of 600 pJ (before any coupling loss); both the experiment and the NLSE simulations show clear pulse broadening and pulse splitting. The temporal full-width half maximum (FWHM) pulse duration increased by a factor of 2, from 2 ps at the input to 4 ps at the output. In addition, clearly defined splitting was noticed at the maximum intensity of the pulse. This is believed to be the result of free-carriers induced by TPA, as the splitting was much less profound when the TPA and FCA was turned off during the NLSE simulation.

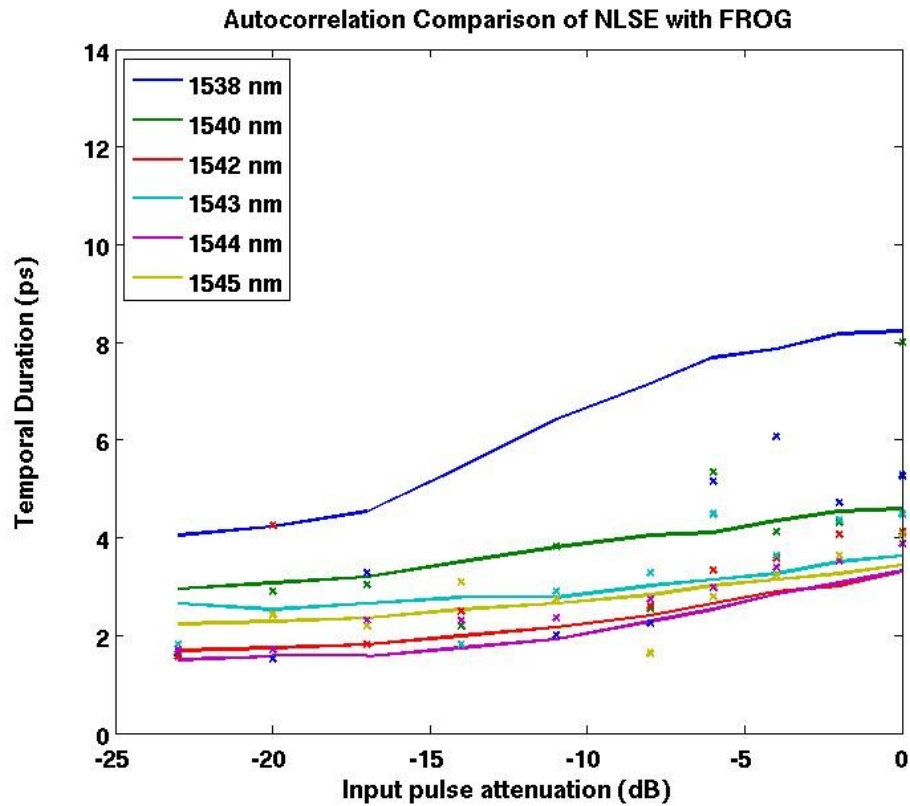

FIG. 4. Comparison of autocorrelation FWHM between the FROG (represented by "X") and the NLSE solver (represented by the solid lines).

Figure 4 shows the comparison of the temporal widths of various wavelengths compared to the intensity of the input pulse energy. The NLSE simulation was run with FROG-measured input pulses, assuming a maximum input pulse of 40 pJ at the entrance of the waveguide; the maximum pulse energy of the mode-locked laser was 600 pJ without any coupling loss. The temporal widths were obtained by convolution and deconvolution of the pulse shapes; a mathematical representation of traditional autocorrelation. In the overall comparison of the temporal widths, the median percent error between the theoretical NLSE and the experimental FROG measurements was 22%. This error can be attributed to not knowing the exact coupling loss in the facets, which can vary the true pulse energy at the input of the waveguide. While no pulse compression was detected at the millimeter length scales, the NLSE model was validated by this experimental effort, thus providing evidence of the feasibility of soliton pulse compression within silicon channel waveguides.

**V. Conclusion**

In conclusion, this experimental and numerical effort has helped develop an understanding of the phenomena of soliton pulse propagation and compression within silicon channel waveguides. The numerical NLSE model has demonstrated that temporal pulse compression can occur with ultrashort and low-powered pulses at centimeter length scales of silicon, even in the presence of TPA and FCA. In addition, FROG measurements of optical pulse propagation in a 4 mm waveguide has helped characterize the initial nonlinear pulse propagation in these waveguides. The experimental results showed the pulse broadening temporally with increasing pulse energies, and even showed signs of pulse splitting as a result of the TPA and FCA. Most importantly, the experimental data validated the NLSE model, as the simulations have closely matched the experimental results. This validation of the NLSE helps to further demonstrate the feasibility of propagating temporal soliton pulses within silicon nanowire-waveguide for the purposes of all-optical data transfer and computing.

**Acknowledgements**

Sources of funding for this effort include Navy Air Systems Command (NAVAIR)-4.0T Chief Technology Officer Organization as an Independent Laboratory In-House Research (ILIR) Basic Research Project (Nonlinear Analysis of Ultrafast Pulses with Modeling and Simulation and Experimentation); a National Science Foundation (NSF) grant (Ultrafast nonlinearities in chip-scale photonic crystals, Award #1102257), the National Science Foundation of China (NSFC) award #61070040 and #60907003, and the Science Mathematics And Research for Transformation (SMART) fellowship. The author's thank Chee Wei Wong, James McMillan, Tingyi Gu, Pin-Chun Hseih, Kishore Padmaraju, Noam Ofir, and the laboratory of Keren Bergman for fruitful discussions, and Mingbin Yu, Guo-Qiang Lo, Dim-Lee Kwong for providing extra silicon waveguide samples in this effort.

# Appendix A: References